\documentclass[12pt]{article}
\usepackage{latexsym}
\usepackage{amssymb}
\usepackage{mathrsfs}
\usepackage{amsmath}
\def\longbar#1{#1\kern-0.7em\raise1.3ex\hbox{{$-$}}}

\def\a{\alpha}   \def\d{\delta}  
   \def\m{\mu}

\def\pa{\partial} 
\def\cb{{\bar C}}
 \def\cB{{\cal B}}  \def\cD{{\cal D}}

  \def\cO{{\cal O}}

\begin{document}
\thispagestyle{empty}

\begin{center} 
  {\large \bf On the renormalisability of gauge invariant extensions
  of the squared gauge potential}\\[5ex] 
  
  {Mboyo Esole\,\footnote{E-Mail:\ esole@lorentz.leidenuniv.nl}
   and Filipe Freire}\,\footnote{E-Mail:\
  freire@lorentz.leidenuniv.nl}\\[2ex]
  Instituut-Lorentz, Universiteit Leiden,\\ P.\,O. Box 9506, 2300 RA Leiden,\\
  The Netherlands\\[4ex]

  {\small \bf Abstract}\\[2ex] 
\begin{minipage}{14cm}
{\small 
We show that gauge invariant extensions of the local functional $\cO =
\frac12\int d^4x A^2$ have long range non localities 
which can only be ``renormalised'' with reference to a specific gauge.
Consequently, there is no gauge independent way of claiming the
perturbative renormalisability of these extensions. 
In particular, they are not renormalisable in the modern sense of
Weinberg and Gomis. Critically, our study does not support the view that
ghost fields play 
an indispensable role in the extension of a local operator into a
non-local one as claimed recently in the literature.\\[4ex]  
} 
\end{minipage} 
\end{center} 
\newpage \pagestyle{plain} \setcounter{page}{1} 


\section{Introduction and summary\label{1}}

In recent years there has been an increasing interest in the
calculation of the square of the gauge potential.
The idea that condensates in Yang-Mills encode
non-perturbative effects has a long history. 
In particular, a condensate of $A^2$ was
considered some years ago \cite{Lavelle:eg} but due to its gauge
dependence it has not been the focus of much investigation.
However, it has been argued in \cite{Gubarev:2000eu} that a non-local 
gauge invariant functional associated to $A^2$ contains
information on topological structures of the Yang-Mills vacuum which
is revealed  by a non vanishing expectation value of the operator.
This scenario is realised in the compact $U(1)$ gauge theory \cite{Polyakov:rs}
where magnetic monopoles condense \cite{Gubarev:2000eu} for large coupling.

The prospect that $A^2$ might indeed carry physically relevant 
information has motivated 
different groups to calculate the expectation value of the local
operator $\cO = \frac12\int d^4x (A_\mu^a)^2$ in covariant gauges. 
There have been both analytic \cite{Kondo:2001nq,Dudal:2003np} 
and numerical studies in the lattice \cite{Boucaud:2001st}.
The working model used has been $SU(2)$ Yang-Mills.
The underlying idea is that the presumed non-local gauge invariant
operator associated with $A^2$ takes a local form in the particular
gauge where the calculations are carried out. In many cases, what are
actually considered are extensions of $\cO$ involving ghost fields
\cite{Kondo:2001nq}.
The introduction of the ghost fields makes it possible to
write the local gauge fixed operator as a gauge fixed BRS invariant
operator. The expected advantages of this
type of approach are the benefit of using the BRS invariance to
guarantee the renormalisability of the composite operator and
the prospect that due to BRST cohomology theorems an observable is
associated to the BRS invariant operator. The latter expectation has
been shown  not to be fulfilled \cite{Esole:2003ke}, more about
which will be discussed below. The question of renormalisability will be
addressed in this letter. 

The renormalisability of a non local gauge invariant extension of a 
local operator (which is given by a local gauge non-invariant expression in a
specific fixed gauge) is addressed here. To summarise our main results,
we present strong arguments that the methods being used to evaluate the
expectation value of $A^2$, or its local extensions, \textit{require a
renormalisation procedure that is slave to a particular gauge fixing}. 
In other words, we can not extend the standard renormalisation procedure, 
based in the introduction of counterterms order by order in perturbation 
theory, outside a unique fixed gauge.  The particular gauge in question 
is the one where the non local extension is expressed by a local functional.
When this occurs, there are no guarantees that the resulting
renormalised expectation value of the non local gauge invariant
extension is gauge independent. This kind of difficulty should not come as a 
surprise. In fact, it would be surprising if
the perturbative methods on which the known renormalisation procedure
is based could capture reliable information on the topological
structure that is expected to be responsible for a non vanishing
expectation value of the gauge invariant extension of $A^2$
\cite{Gubarev:2000eu}.

In our study we illustrate the properties of gauge invariant extensions 
of local functionals.  We aim at clarifying, via specific examples, 
the relation between a functional which is local in a particular gauge 
(but not necessarily gauge invariant), and its gauge invariant extension 
(which is not necessarily local).  We show that the non localities found
are not perturbatively local because they can not be expressed in terms of 
an infinite derivative expansion. We believe that the implications of
this observation have not been clearly emphasised in the literature,
as attested by the absence of any debate about it in recent works.
It is precisely these dangerous infrared modes that make it hard to
define a gauge independent renormalisation for the gauge invariant
extensions of local functionals. This observation supports the
remark in \cite{Gubarev:2000eu} that the expectation value receives
important contributions from both large and small distances. 
Our arguments on renormalisability are based on the notion of
renormalisation in the modern sense \cite{Gomis:1995jp} which relies
on BRST cohomology theorems. The BRST terminology will therefore be
frequently used here, even though it is not always necessary.

The expectation value of the extensions of $A^2$ can only be claimed to be
``renormalisable'' in the particular gauge where they have a local expression. 
If we try to define renormalisability in any gauge by going back to
the gauge where the functional takes a local form, we become tied to
this gauge and renormalisability is no longer a gauge independent
property. Implicitly, this is what has been done to date in the
literature \cite{Kondo:2001nq,Dudal:2003np,Boucaud:2001st}.  We
emphasise the distinction between this situation and, for example, the
case of the standard model where renormalisability can be guaranteed
from gauge to gauge.

The non locality of a gauge invariant extension of $\cO$ is
unavoidable if the operator is to be associated to an observable.
In a recent analysis \cite{Esole:2003ke} we have
studied the operator $\cO$ in the powerful context of the antifield
Batalin-Vilkovisky formalism \cite{Batalin} using local BRST extension
and deformation techniques \cite{HenneauxReport}. 
By analysing $\cO$ with a ghost sector we have shown that ghost
condensates are an artifact of gauge fixed actions. 
A by-product of this analysis was the observation that there
is no local observable associated with an on-shell BRS invariant
mass dimension two local functional in $SU(N)$ Yang-Mills theories.
This observation has important consequences.

On the one hand, it illustrates that gauge-fixed BRS invariance and gauge 
invariance \cite{HenneauxGF1} are not always equivalent.   
In this respect it is important to realise that the relation between
classical observables and gauge-fixed BRST cohomology is not a
straightforward one, unlike the case for the gauge independent BRST cohomology.
Only in this last case are we guaranteed the existence
of a one-to-one correspondence between classical observables and
elements of the BRST cohomology at zero ghost number for both local
and non-local functionals.  For \textit{non-local}
functionals the correspondence is one-to-one even for the gauge-fixed
case, while for local functionals extra conditions (see discussion at
end of Sect.~\ref{5}) which are not fulfilled by $\cO$ are required
\cite{HenneauxGF1}. 

On the other hand, it indicates that
the only viable extensions of $\cO$ ought to be non local.
The possibility of associating a non-local observable to a mass
dimension two operator has since been exploited by Kondo \cite{Kondo:2003uq} 
in order to argue for a physical meaning to $\cO$ in the Abelian gauge
theory. In his study the author makes use of gauge-fixed
cohomology to derive the possible physical interpretation of $\cO$. 
Some subtle aspects of Kondo's arguments require re-evaluation.
In general, his line of reasoning would make any on-shell
BRS invariant operator a candidate for a physical observable.
If we have in mind that one of the features of gauge theories is their
constrained structure this is too generic.

We state here three key observations that conflict with Kondo's view. 
Firstly, the vacuum expectation value of $\cO$ is only
equivalent to that of its gauge invariant extension in a particular
gauge and, therefore, this equivalence cannot be the origin for gauge
independent statements.
Secondly, ghost fields are not essential to extend a local gauge variant
operator into a non local invariant one. Thirdly, the way a gauge
condition is implemented has an effect on the BRST transformations.
All these issues will be discussed throughout this letter along side
the central question of renormalisation.

To avoid the technical difficulties involved in Yang-Mills theories
the explicit examples we use in this letter are in Maxwell's theory.
However, all the properties of gauge invariant extensions we illustrate
are generic, and also apply to the non-Abelian case.
In Sect.~2 we present a general discussion of gauge invariant
extensions. This is followed by an explicit study of non-local
extensions of $\cO$ in Maxwell's theory for arbitrary linear gauges
in Sect.~3. We illustrate that the properties of gauge
invariant extensions are related to those of the gauge in which $\cO$ 
is initially specified, and emphasise the fact that locality is often not
preserved by the extension procedure.
In Sect.~4 we discuss the two standard ways of implementing gauge
fixing because of the importance of specifying the gauge from which
the extension is constructed.
In Sect.~5 we analyse the gauge dependent nature of the relation
between the expectation values of $\cO$ and its gauge invariant
extension. Various subtle issues concerning the renormalisation of
non-local gauge invariant extensions are discussed in Sect.~6.
In Sect.~7 we present a final discussion on our
analysis. 

\section{Gauge invariant extension\label{2}}

Consider a local functional $\cO$ and a fixed gauge.
The latter is specified by the gauge fixing fermion $\Psi$ following
the prescription where the gauge fixing plus ghost sector of the
action is $\int s\Psi$, with $s$ the BRST operator.
The functional $\cO$ can always be extended off the gauge $\Psi$ in a
gauge invariant way.  The resulting functional, the \textit{gauge invariant
extension}, which we denote by
${\cO}^{\uparrow\Psi}$, is by construction strongly gauge invariant
\cite{HenneauxBook}.
Unless $\cO$ is itself gauge invariant the relation between $\cO$ and
${\cO}^{\uparrow\Psi}$ depends on the specified gauge $\Psi$,
therefore we keep $\Psi$ as an upper script in the extension as a reminder.
An important gauge dependent identity that follows from the
construction of the extension ${\cO}^{\uparrow\Psi}$ is the equality
\begin{equation}
  \label{eq:equal-expectation-values}
  \langle \cO \rangle_{\Psi} =   \langle {\cO}^{\uparrow\Psi}
  \rangle_{\Psi},
\end{equation}
where $\langle\,\cdot\,\rangle_{\Psi}$ denotes the expectation value
evaluated in the specific gauge $\Psi$. We will return to
\eqref{eq:equal-expectation-values} at a later stage.

Note that though the gauge invariant extension is not necessarily
unique for a given functional $\cO$ in a gauge $\Psi$, two different extensions
${\cO}^{\uparrow\Psi}$ and ${\cO'}^{\uparrow\Psi}$ always have the same
expectation value.  This follows from the fact that the ambiguity is
proportional to terms that vanish modulo the equations of motion of the
\textit{gauge invariant} action or $s$-exact terms.

In the examples discussed in this letter the extensions can be
computed using only fields and ghosts. Moreover, the final explicit 
expressions for the gauge invariant extensions can be written without 
ghosts. Therefore, contrary to \cite{Kondo:2003uq} \textit{we do not find 
evidence that ghost condensates are necessary to convert local
operators into non-local ones}. 

The gauge $\Psi$ which we shall call the ``\textit{base gauge}''
has an important role in determining which properties 
and symmetries of $\cO$ are carried along to $\cO^{\uparrow\Psi}$.
For example, if we extend a covariant operator $\cO$ from a non covariant
base gauge, the resulting extension is not expected to be covariant.
This will be illustrated below.
Independently of the base gauge, another property of $\cO$ that the
gauge invariant extension does not normally preserve is locality.
Indeed, the extension is in general non local
unless $\cO$ is local and gauge invariant modulo the
equations of motion of the \textit{gauge invariant} action.

The construction of a gauge invariant extensions of a functional
starting from a base gauge is very generic and
in this sense it is always possible to associate a gauge invariant
quantity to any $\cO$. It should however, be emphasised that the
methods used here only apply for extensions on a local patch because
it is still possible to have obstructions due to the topological
structure of the configuration space. As long as one works in
perturbation theory these obstructions are avoided.

\section{The $A_\mu^2$ functional in the Maxwell theory\label{3}}

Consider the free Abelian gauge theory in four dimensions and let
$\cO$ denote the gauge dependent mass dimension two local functional
\begin{equation}\label{O}
  \cO=\frac12\int\!d^4x\ A_\mu^2\,.
\end{equation}
From the BRS transformation of the gauge potential,
$sA_\mu=\pa_\mu C$, we have that the variation of $\cO$ is given by
  \begin{equation}
    \label{mass-transform}
    s\cO=\int\!d^4x\  A_\m\,\pa^\m C=-\int\!d^4x\ \partial\!\cdot\!A\ C.
  \end{equation}
It follows from \eqref{mass-transform} and the discussion in
\cite{Esole:2003ke} that $\cO$ can not be added to the action as a
mass term without effectively changing the physical content of the theory.  
The functional $\cO$ is used in this letter
to construct gauge invariant extensions from various base gauges.
This will provide us with explicit examples to study some general
properties of these extensions.

We start by computing the gauge invariant extension of
$\cO$ for a general linear gauge $\Psi_\ell$ as the gauge base.
The gauge condition is given by
  \begin{equation}\label{linearGauge}
  \ell\!\cdot\!A \equiv \ell^\m A_\m =0\,,
  \end{equation}
where $\ell^\mu$ is an $A_\mu$ independent linear operator. Three
familiar choices will be considered here,
\begin{eqnarray}
  \ell^\m\!\!\!&=&\!\!\!\pa^\mu,\hspace{.2cm} 
  \mbox{Lorentz gauge ($\pa\!\cdot\!A=0$)},\label{lorentz}\\ 
    \ell^\m\!\!\!&=&\!\!\!n^\mu, \hspace{.2cm} 
    \mbox{general axial gauge ($n\!\cdot\!A=0$),}\label{axial}\\ 
       \ell^\m\!\!\!&=&\!\!\!\pa^\m-\d^\m_0 \pa^0, \hspace{.2cm} 
       \mbox{Coulomb gauge ($\vec\partial \cdot \vec A=0$)}\label{coulomb},
\end{eqnarray}
where $n^\m$ is a fixed 4-vector.
The idea behind the calculation of $\cO^{\uparrow\Psi_\ell}$ is 
very simple. Consider the infinitesimal variation of $\cO$ along
the gauge orbit when the potential is shifted away from the
base gauge. Then look at how to modify $\cO$ so
it is parallel transported along the gauge orbit.
Here for later convenience we consider the variations of the potential to be
of the form of a BRS transformation where a ghost field appears at the
place of the infinitesimal variation of the gauge parameter.
 
By applying the linear operator $\ell_\mu$ on both sides of 
$s A_\mu = \partial_\mu C$ we obtain a non-local expression for the
ghost field in terms of the gauge potential,
\begin{equation}
C=s\,\Big(\frac{\ell\!\cdot\!A}{\ell\!\cdot\!\pa}\Big) .\label{eq:c-transform}
\end{equation}
For example, in the Lorentz gauge, 
$\frac{\ell\cdot A}{\ell\cdot \pa}=\frac{\pa\cdot
    A}{\square} = -\int 
    d^4k\,\frac{k^\mu\tilde{A}_\m(k)}{4\pi^2k^2}\,e^{ikx}+h.c.$,
in the usual representation using Fourier transforms and
distribution theory.

Using \eqref{eq:c-transform} it is now straightforward to determine
$\cO^{\uparrow\Psi_\ell}$.  From
\eqref{mass-transform} and \eqref{eq:c-transform} we have
\begin{eqnarray}
   s\cO &=& -\int d^4x\ \pa\!\cdot\!A\  s\,\Big(\frac{\ell\!\cdot\!A}
   {\ell\!\cdot\!\pa}\Big)\nonumber\\
   &=& -\int d^4x\ \Big(s\,\Big(\frac{\ell\!\cdot\!A}
   {\ell\!\cdot\!\pa}\ \pa\!\cdot\!A\Big) -
   \frac{\ell\!\cdot\!A}{\ell\!\cdot\!\pa}\ \square C\Big) \nonumber\\
   &=&-s\int d^4x\! \left(\frac{\ell\!\cdot\!A}{\ell\!\cdot\!\pa}\
   \pa\!\cdot\!A 
   -\frac12\ \frac{\ell\!\cdot\!A}{\ell\!\cdot\!\pa}\ 
   \square\frac{\ell\!\cdot\!A}{\ell\!\cdot\!\pa}\right).
 \end{eqnarray}
and we arrive at the BRS invariant extension
\begin{equation}
{\cO}^{\uparrow\Psi_{\ell}}=\frac12 \int d^4x\! \left(A^2
   +2\ \frac{\ell\!\cdot\!A}{\ell\!\cdot\!\pa}\ \pa\!\cdot\!A
   - \frac{\ell\!\cdot\!A}{\ell\!\cdot\!\pa}\ 
   \square\frac{\ell\!\cdot\!A}{\ell\!\cdot\!\pa}\right),
  \label{eq:extension}
\end{equation}
which is also strongly gauge invariant in any local patch.
The functional in \eqref{eq:extension}  can be naturally
identified as the gauge invariant extension of $\cO$ in a linear gauge 
in the sense that
\begin{equation}
  \label{eq:descend}
  {\cO}^{\uparrow\Psi_{\ell}}\Big\vert_{\ell\cdot A=0} \!\!= \cO\,.
\end{equation}
We can see from \eqref{eq:extension} that the extension depends on
the base gauge has expected. In particular, for the Lorentz gauge
we have
 \begin{equation}
 {\cO}^{\uparrow\Psi_{\mathrm{L}}} =
   \frac12\int d^4x\,\bigg(A^2
   + \frac{\pa\!\cdot\!A}{\square}\ \pa\!\cdot\!A\bigg),\label{ext1}
 \end{equation}
which is clearly non local though $\cO$ is local. We make the
important observation that the non locality in \eqref{ext1} can not be
expanded in a Taylor series. This property will be central to the
discussion in Sect.~\ref{5}. Similar types of non locality occur for
extensions from other base gauges. For the axial gauge we have
 \begin{equation}
        {\cO}^{\uparrow\Psi_{\mathrm{A}}} =
        \frac12 \int d^4x\,\bigg(A^2
        +2\ \frac{n\!\cdot\!A}{n\!\cdot\!\pa}\ \pa\!\cdot\!A - 
   \frac{n\!\cdot\!A}{n\!\cdot\!\pa}\ \square\, 
        \frac{n\!\cdot\!A}{n\!\cdot\!\pa}\bigg),\label{ext2}
 \end{equation}
and for the Coulomb gauge
 \begin{equation}
             {\cO}^{\uparrow\Psi_{\mathrm{C}}} =
             \frac12\int d^4x\,\bigg(A^2+2\ 
             \frac{\vec\pa \cdot \vec A}{{\vec\pa}\,{}^2}\ \partial\cdot A
             - \frac{\vec\pa\cdot \vec A}{{\vec\pa}\,{}^2}\ \square\,
             \frac{\vec\pa\cdot \vec A}{{\vec\pa}\,{}^2}\bigg).\label{ext3}
 \end{equation}
Another property of the extensions concerns the effect the base gauge
has upon the symmetries of $\cO$.  In the above examples 
we always started with a covariant operator
but only the extension \eqref{ext1} preserves covariance.  The covariance
in \eqref{ext2} and \eqref{ext3} is lost in the process of extending 
$\cO$ away from a non covariant base gauge.
 
\section{Gauge fixing implementation\label{3b}}

By definition the gauge invariance extension requires the choice of
a specific base gauge as a starting point. It is therefore interesting
to analyse how the extension might be affected by the gauge fixing procedure.
When BRS techniques are used there are two standard implementations to fix
the gauge, the delta function and the Gaussian average.

So far we have implemented the gauge fixing by
requiring a gauge condition to be explicitly satisfied, \eqref{linearGauge}.
In a path integral representation this corresponds to implementing the
gauge condition via a delta function. As an example, for the Lorentz
gauge, \eqref{lorentz}, the 
gauge fermion is $\Psi^{(\delta)}_\mathrm{L} = \cb\,(\pa \cdot A)$. 
The corresponding gauge fixing and ghost sector of the action is 
$\int s\Psi^{(\delta)}_\mathrm{L} = \int b\,(\pa \cdot A) - \cb\,\square\,C$,
where $b$ is the auxiliary Nakanishi-Laudrup scalar. It follows
\begin{equation}
  \label{eq:deltacondition}
  \int {\cal{D}}[A_\mu,\cb,C,b]\exp{i(S+{\textstyle\int}s
 \Psi^{(\delta)}_\mathrm{L})}=
  \int{\cal{D}}[A_\mu]\,\det\square\ \delta(\pa\!\cdot\!A)\exp{iS}.
\end{equation}

We consider now the other common way of implementing gauge
fixing: Gaussian averaging of the gauge condition. This
implementation is equivalent to the delta function one at the
level of the gauge independent BRST antifield formalism.
However, the Gaussian averaging is the appropriate one to introduce
the gauge-fixed BRST cohomology and analyse its relation to the
off-shell gauge invariant formulation \cite{HenneauxGF1}.
As we will see, this implementation is more general as it contains the
previous in a specified limit.

The gauge fermion that implements the Lorentz condition by Gaussian
averaging is $\Psi^{(\mathrm{Gauss})}_\mathrm{L} = \cb\,(\pa \cdot A -
\frac\alpha2\,b)$, where $\a$ is the gauge fixing parameter.
It follows from the gauge-fixed action that the equation of motion for the
auxiliary field is $b=\frac1\alpha\,\pa\!\cdot\!A$.  The
on-shell gauge-fixed BRS transformations are obtained after
integrating over $b$ i.e.~by implementing the $b$ equation of motion. 
As an example, in the Lorentz gauge the off-shell BRS transformations
$s\cb=b$, $sb=0$ become $s\cb=\frac1\alpha\,\pa\!\cdot\!A$, $sb=0$.
For a delta function implementation of a gauge fixing condition these
on-shell transformations can not be derived because $b$ only enters
linearly in the gauge-fixed action. 

The path integral representation for the Gaussian averaging
of the Lorentz condition is
\begin{eqnarray}
  \label{eq:Gaussianaver}
  \int {\cal{D}}[A_\mu,\cb,C,b]\exp{i(S+{\textstyle\int}s
  \Psi^{(\mathrm{Gauss})}_\mathrm{L})}=\hspace{4cm}\nonumber\\
  \int{\cal{D}}[A_\mu]\,\det\square\ 
  \exp{i(S+{\textstyle{\frac1{2\alpha}\int}}(\pa\!\cdot\!A)^2)}.
\end{eqnarray}
In the limit $\alpha\to0$ the delta function implementation is
recovered.

\section{The expectation value of ${\cO}^{\uparrow\Psi_{\ell}}$\label{4}}

Let us consider for the moment the
Gaussian average gauge fixing implementation. The phase space needs to be 
extended to include the antighost and the auxiliary field and the
general linear gauge corresponds to the condition $\a b = \ell\cdot A$. 
Then, following an analogous approach to the one of Sect.~\ref{3}, the gauge
invariant extension of $\cO$ for a base gauge specified by this condition is
\begin{equation}
   {\cO}^{\uparrow\Psi_{\ell}}=\frac12 \int d^4x\left( A^2
   +2\ \frac{\ell(A,b;\a)}{\ell\!\cdot\!\pa}\ \pa\!\cdot\!A
   - \frac{\ell(A,b;\a)}{\ell\!\cdot\!\pa}\ 
   \square\frac{\ell(A,b;\a)}{\ell\!\cdot\!\pa}\right),\label{eq:extension2} 
\end{equation}
with $\ell(A,b;\a) = \ell\cdot A - \a b$. Note 
that the $\a = 0$ choice corresponds to the extension \eqref{eq:extension}.
Moreover, the right-hand sides of \eqref{eq:extension} and
\eqref{eq:extension2} are equal up to $s$-exact terms as 
they only differ by terms involving the auxiliary field.
They correspond therefore to the same gauge invariant functionals and
we will be using the simpler form \eqref{eq:extension}. 
At this point, it is also important to note that if we modify the
integrand of the functional by adding a ghost sector, $\frac12 A^2 \to
\frac12 A^2 - \a C\bar C$, and compute the gauge invariant extension,
the resulting extension will only differ by an $s$-exact term, 
$-\a s\int \bar C\, (\ell\!\cdot\!\pa)^{-1}\ell(A,b;\a)$. 
It should be remarked that this is not a specific property
for this functional $\cO$, as it follows alone from the fact that
the non-local gauge invariant extension of any term involving $C\bar
C$, or for this purpose any other auxiliary fields, will
always give trivial elements on the cohomology of $s$.
Moreover, as it has been shown in \cite{Esole:2003ke} $C\bar C$ does
not have local extensions. 

For a general linear gauge, 
 $\Psi_\ell=\ell\!\cdot\!A-\frac12\a b$, the equation of motion of
 $b$ reduces to 
 \begin{equation}
     b=\frac{1}{\a}\ \ell\!\cdot\!A\,,\label{eq:b-eom}
 \end{equation}
 and therefore $\ell(A,b;\a)=0$. From \eqref{eq:b-eom} it follows
 that the on-shell gauge-fixed BRS symmetry in the linear gauge is expressed by
\begin{equation}\label{onshell}
    s_{\Psi_\ell} \cb=\frac1\a\ \ell\!\cdot\!A\,,
\end{equation}
where $s_{\Psi_\ell}$ is the corresponding gauge-fixed BRS operator.
At this level the equation of motion has been already
implemented or equivalently, the $b$ field has been integrated over.
By taking \eqref{onshell} into account, the non-local terms on the
right-hand side of \eqref{eq:extension} can be expressed as
\begin{eqnarray}\label{eq:term1}
   \frac{\ell\!\cdot\!A}{\ell\!\cdot\!\pa}\ \pa\!\cdot\!A
   =\a\cb\frac{ \square}{\ell\!\cdot\!\pa}\,C +
   s_{\Psi_\ell}\Big(\frac{ \a\cb}{\ell\!\cdot\!\pa}\,\pa\!\cdot\!A\Big)\,,
\end{eqnarray}
and
\begin{eqnarray}\label{eq:term2}
    \frac{\ell\!\cdot\!A}{\ell\!\cdot\!\pa}\ 
    \square\frac{\ell\!\cdot\!A}{\ell\!\cdot\!\pa}
    =\a\cb\frac{\square}{\ell\!\cdot\!\pa}\,C +
    s_{\Psi_\ell}\Big(\frac{\a\cb}{\ell\!\cdot\!\pa}\ 
    \square\frac{\ell\!\cdot\!A}
    {\ell\!\cdot\!\pa}\Big)\,.
\end{eqnarray}
Inserting (\ref{eq:term1}-\ref{eq:term2}) into
\eqref{eq:extension} gives the explicit
relation between $\cO$ and ${\cO}^{\uparrow\Psi_{\ell}}$,
\begin{eqnarray}\label{relation}
    {\cO}^{\uparrow\Psi_{\ell}}=\cO + \frac\a2 \int 
    d^4x\ \cb\frac{\square}{\ell\!\cdot\!\pa}\,C
    + s_{\Psi_\ell}\cB\,,
\end{eqnarray}
where $\cB = \a\int d^4x \frac{\bar C}{\ell\cdot\pa}(\pa\!\cdot\!A -
\frac12 \square\frac{\ell\cdot A}{\ell\cdot\pa})$ is a functional of
the fields and ghosts. If we use 
\eqref{eq:extension2} instead of \eqref{eq:extension} as the
expression for the gauge invariant extension, the
relation~\eqref{relation} remains valid but $\cB$ is different. 
From this equation the expectation value  of $\cO$
is in general not equal to ${\cO}^{\uparrow\Psi_{\ell}}$ and an equality
is only guaranteed in the gauge $\Psi_{\ell}$. Therefore, no gauge
independent statement can be made between
$\langle{\cO}^{\uparrow\Psi_{\ell}}\rangle$
and $\langle\cO\rangle$.

In order to clarify this point we take a closer look at the last two
terms in the right-hand side of \eqref{relation}.  Let us first recall
the standard principle behind Ward identities. Consider 
$\delta$ to denote a classical symmetry of the action. Then for any
functional ${\cal F}$ we have $\langle\delta{\cal F}\rangle = 0$.
As far as $s_{\Psi_\ell}$ is concerned,  as this refers solely to a
symmetry of the gauge-fixed action for $\Psi=\Psi_\ell$ we can only state that 
$\langle s_{\Psi_\ell}{\cal F}\rangle_{\Psi_\ell} = 0$.
Therefore, $\langle s_{\Psi_\ell}\cB \rangle_{\Psi_\ell}=0$,
but in general $\langle s_{\Psi_\ell}\cB \rangle_{\Psi}\neq0$.

Next, consider the identity
\begin{equation}\label{id}
s_{\Psi_{\ell}}\Big(\bar C\,\frac{\pa\!\cdot\!A}
{\ell\!\cdot\!\pa}\Big) = \frac1\a\,\ell\!\cdot\!A\ \frac{\pa\!\cdot\!A}
{\ell\!\cdot\!\pa} - \bar C \frac{\square}{\ell\!\cdot\!\pa}C\,.
\end{equation}
The expectation value of the left-hand side vanishes in the gauge
$\Psi_{\ell}$. The same also applies to the first term on the right-hand
side. This follows from the off-shell identity 
$\langle b\, \frac{\pa\cdot A}{\ell\cdot\pa} \rangle = 
\langle s_{\mathrm{aux}}( \bar C\, \frac{\pa\cdot A}{\ell\cdot\pa})
\rangle=0$, where $s_{\mathrm{aux}}=s$ when acting on $\bar C$
and $b$ and gives zero on the other fields. Because the $b$ field only
enters linearly in this identity $\langle\ell\!\cdot\!A\, \frac{\pa\cdot A}
{\ell\cdot\pa}\rangle_{\Psi_\ell}=0$. Therefore, we have from \eqref{id}
that in the general linear gauge
$\langle \cb\frac{\square}{\ell\cdot\pa}\,C \rangle_{\Psi_\ell}=0$
for the Maxwell theory. We then arrive at the \textit{de facto} gauge
dependent equality 
\begin{equation}
  \label{eq:exp-value}
  \langle \cO \rangle_{\Psi_\ell} = 
  \langle {\cO}^{\uparrow\Psi_{\ell}} \rangle_{\Psi_\ell}
\end{equation}
as expected from \eqref{eq:equal-expectation-values} which is a direct
consequence of the construction of gauge invariant extensions. 

\section{On the renormalisation of non-local functionals\label{5}}

In this section we discuss the perturbative renormalisability of the 
operator $\cO$ in the \textit{modern sense} as introduced by Gomis 
and Weinberg \cite{Gomis:1995jp}.  This criterion extends the Dyson
one by allowing terms that are not power counting renormalisable.
A theory is said to be renormalisable in the modern sense, if the
symmetries of the bare action provide constraints that are sufficient
to eliminate all the infinities. The symmetries of the bare action
are encoded in the BRST symmetry of the gauge invariant action in 
the antifield formalism which is gauge independent.
Gauge independent statements on the renormalisability of a 
given gauge theory are made possible by the close link 
between this renormalisation criterion and the cohomology of the
BRST transformations generated by the action.
Well established local BRST cohomology theorems
\cite{Barnich:ve,Barnich:mt} provide the criteria to identify all the
possible local counterterms. Contrary to the power-counting
renormalisation criterion, there is no limit on the mass dimension of
the allowed terms in the bare action. Therefore,
an infinite number of counterterms are viable.

A sufficient condition for the renormalisability of the theory is the
existence of an independent coupling in the action for each
non-trivial element of the BRST cohomology.  It is important to note 
that we can add any local term to the action compatible with the
theory symmetries. In particular, we can add a non-local term in the
form of an infinite number of derivative terms. It is still possible in 
this case to have a theory that is renormalisable in the modern sense
because each derivative term is local, as required by the Quantum
Action Principle \cite{QAP}. An example occurs when the
non locality enters through terms
of the form $(\square+m^2)^{-1}$ which can be expressed
as an infinite sum of local terms 
$\sum_{n=0}^\infty m^{-(2n+2)}(-\square)^n$, as long as $m\neq0$.
In this sense, even the Wilson loop is a perturbatively local quantity
because it can be expressed in terms of an infinite series of local
terms \cite{Shifman:ui}.

Here we are interested, in particular, in the renormalisability of 
a non-local gauge invariant functional like the extensions
(\ref{ext1}-\ref{ext3}).  The non locality in these extensions can not
be expressed in terms of an infinite series of local terms. From the 
discussion in the previous paragraph we conclude that there is no gauge
independent way in which $\cO^{\uparrow\Psi}$ is renormalisable in the
modern sense. Because of the formal relation
$\langle \cO^{\uparrow\Psi} \rangle = -i\frac{\delta}{\delta J}\int\cD\phi 
\exp(i S[\phi] + iJ\cO^{\uparrow\Psi})\vert_{J=0}$, as far as the
role of the non locality is concerned, the non
renormalisability of $\cO^{\uparrow\Psi}$ can be inferred from that of
theory where the functional $\cO$ is coupled to a source $J$ and
inserted to the action.

This, of course, is not in contradiction with the fact that 
an extension $\cO^{\uparrow\Psi}$ can be perturbatively 
renormalisable in the base gauge where it takes a local form.
What happens in this particular case is that \textit{the ``local'' counterterms
that make the functional renormalisable in this gauge can not be
expressed as a series of local terms in other gauges}.
We used ``local'' in the last sentence, to emphasise that for the
consistency of the renormalisation procedure, locality should not be
restricted to a particular gauge. However, this is not guaranteed in
the present examples and therefore the Quantum Action Principle,
which requires all the counterterm to be local is not ensured for other
gauges. 

It is interesting to see that starting from a gauge where $\cO$ is
multiplicatively renormalisable, the covariant gauge with $\a=0$,
that the renormalisability can not be ``extended'' to
$\cO^{\uparrow\Psi}$ \textit{without having to redefine the standard
renormalisation procedure}.
A local functional can be associated to a gauge invariant quantity if
it fulfills the two following conditions \cite{HenneauxGF1}: 
\begin{enumerate}
\item it must be on-shell BRS invariant;
\item it must not break the nilpotency of the BRS symmetry when it is
  added to the gauge-fixed action. 
\end{enumerate}
In order for these conditions to hold, one must use the Gaussian averaging
implementation of the gauge fixing. With a delta function
implementation the first condition is not even satisfied (for non
gauge invariant functionals). For example, consider $\cO$ in the gauge
$\pa\cdot A=0$. From \eqref{mass-transform} 
we see that the BRS variation of $\cO$ can not vanish on-shell
as the equations of motion in the gauge $\Psi^{(\d)}_\mathrm{L}$
are $\pa^\mu F_{\mu\nu} + \pa_\nu b = \square C=\square\bar C=0$.
The only way to have $s\cO=0$ is to set $\pa\cdot A=0$, i.e.~impose
the gauge condition ``by hand''. The subtle distinction on the
implications of these different gauge fixing implementations was not clearly
distinguished in \cite{Kondo:2003uq}. 

The first condition can be satisfied by considering the modified operator 
$A_0=\cO - \a \int d^4x\,C\bar C$ in the gauge fixed by 
$\Psi^{(\mathrm{Gauss})}_\mathrm{L}$, where now $sA_0=0$ modulo the equation of
motion of $b$, $b=\frac1\a\,\pa\!\cdot\!A$. However, this operator
does not fulfill the second condition because $s_{\Psi}^2 \bar C$ 
no longer vanishes on-shell.

\section{Discussion\label{6}}

In this letter we have analysed the properties of non-local
gauge invariant functionals by studying some simple examples.
We used general extension methods to compute gauge invariant functionals
$\cO^{\uparrow\Psi}$ by transporting a local functional $\cO$ defined
in a specified base gauge $\Psi$ away from this gauge. We have looked
explicitly at gauge invariant extensions for the mass dimension two
functional $\cO = \frac12\int d^4x A_\mu^2$ in Maxwell's theory.
From our previous analysis \cite{Esole:2003ke} in Yang-Mills theories
it follows that these extensions have to be non local.

The non-local functionals encountered in our computation of gauge invariant 
extensions from general linear gauges \eqref{eq:extension} are not of
the type that can be handled perturbatively. The non localities result
from long range fluctuations that can not be renormalised by perturbative 
methods even when one calls for an infinite set of local counterterms.
In this sense, the functionals in our examples are not renormalisable
in the modern sense.
The situation for gauge invariant extensions in Yang-Mills theories
for a functional of the form $\cO = \frac12\int d^dx\,(A_\mu^a)^2$
is even more problematic. Besides having to deal with the same type of
long range non localities the various non localities interact in
a non-polynomial way.

We are well used to the idea that we need to fix the gauge in
perturbation theory. However, when dealing with (perturbatively) local
functionals we know that by changing the gauge all the counterterms
remain local in accordance to the Quantum Action Principle.
For the non-local gauge invariant extensions it is only in the base
gauge that the counterterms are guaranteed to be local.  

Therefore, renormalisability can only be claimed with reference to one
particular gauge \cite{Kondo:2001nq,Dudal:2003np,Boucaud:2001st}.
In other words, the only known way to make gauge invariant extensions 
renormalisable is by redefining renormalisability by
\textit{construction} in the base gauge of the extension, 
i.e.~$\langle\cO^{\uparrow\Psi}\rangle_{\Psi'} = 
\langle\cO\rangle_{\Psi}$ for any gauge $\Psi'$.
In this way there is a clear prescription to claim 
$\langle\cO^{\uparrow\Psi}\rangle$ to be ``renormalisable'' -- however
the procedure is gauge dependent.  As a result, the theory only lives in
one gauge with reference to which any calculation of quantities
involving insertions of $\cO^{\uparrow\Psi}$ is possible.
A well-known example of this situation is illustrated by the
Curci-Ferrari model \cite{Curci:bt,Brandt}.

This makes unclear the status of the physical relevance of
$\cO^{\uparrow\Psi}$ although it is gauge invariant. At the very
least, a necessary condition for the relevance of the constructed 
gauge invariance of $\cO^{\uparrow\Psi}$ is the existence of a
renormalisation procedure without reference to a specific gauge.

In addition, by constructing non-local gauge invariant extensions
from local functionals there is, in principle, an endless
line of candidates for observables.  Each can be made local in
a particular ``proper'' gauge, as our examples illustrate. The extension
procedure is too generic and does not provide \textit{by itself},
and without the constraint of perturbative locality \cite{Esole:2003ke}, a
strong claim to support the physical relevance 
for a functional that it is not gauge invariant.

We conclude that a well defined meaning of such functionals without
reference to the gauge where they are local and polynomial is missing.  
The current methods used to
compute renormalised functionals require assumptions that are only
known to be fulfilled by perturbatively local functionals but not by
the type of non-local functionals found in the present letter. 
The development of the non-perturbative methods to renormalise
non-local functionals in a gauge independent manner 
without the constraint of the Quantum Action
Principle might help to improve our understanding about the relevance
of gauge invariant extensions which are not perturbatively local.\\[.4cm]

\noindent
\textbf{\Large{Acknowledgments}}\\

\noindent
FF thanks Jan Pawlowski for useful discussions. ME thanks Glenn
Barnich for useful references.
The work of FF is supported by FOM.

\end{document}